\documentclass[pre,twocolumn,groupedaddress,showpacs,floatfix]{revtex4}
\usepackage{graphicx}
\usepackage{dcolumn}
\usepackage{amsmath}
\usepackage{bm}
\begin{document}
\title{Defect structures and torque on an elongated colloidal particle immersed in a liquid crystal host}
\author{Denis Andrienko}
\altaffiliation{Max Planck Institute for Polymer Research,
Ackermannweg 10, 55128 Mainz, Germany}
\author{Michael P. Allen}
\altaffiliation{Center for Scientific Computing, 
University of Warwick, Coventry, United Kingdom}
\affiliation{H. H. Wills Physics Laboratory, University of Bristol,
Royal Fort, Tyndall Avenue, Bristol BS8 1TL, United Kingdom}
\author{Gregor Ska{\v c}ej}
\author{Slobodan {\v Z}umer}
\affiliation{Physics Department, University of Ljubljana, Jadranska 19, 
SI-1000 Ljubljana, Slovenia}
\date{\today}
\begin{abstract}
 Combining molecular dynamics and Monte Carlo simulation we study 
 defect structures around an elongated colloidal particle embedded in a nematic
 liquid crystal host. By studying nematic ordering near the particle 
 and the disclination core region we are able to examine the 
 defect core structure and the difference between two simulation techniques. 
 In addition, we also study the torque on a particle tilted with respect to
 the director, and modification of this torque when the particle is close to 
 the cell wall.
\end{abstract}
\maketitle

\section{Introduction}

Colloidal dispersions of small particles in nematic liquid
crystals are a novel type of soft matter. 
Topological defects \cite{lubensky.tc:1998.a,stark.h:1999} and 
additional long-range forces between the colloidal particles 
\cite{lev.bi:1999.a,borstnik.a:1999.a,borstnik.a:2000.a} 
are immediate consequences of the orientational 
ordering of the liquid crystal molecules.
The nematic-induced interparticle interaction brings a new range of 
effects to the system: supermolecular structures
\cite{poulin.p:1997.a,poulin.p:1998.a}, cellular structures
\cite{anderson.vj:2001.a, anderson.vj:2001.b}, and even a soft
solid \cite{meeker.sp:2000.a} can be observed. Colloidal
dispersions in liquid crystals also have a wide variety of
potential applications \cite{russel.wb:1989.a}.

The subject of this paper is the liquid crystal ordering and equilibrium 
orientation of an elongated solid particle inside a uniformly aligned 
nematic liquid crystal. On the list of problems one has to clarify are: 
  (1) the nematic ordering around the particle (including possible 
      topological defects); 
  (2) the type and strength of the orientational coupling between 
      the particle and its aligned molecular environment;
  (3) the effect of confinement on the orientational ordering of 
      the particle, i.e. the equilibrium orientation of the 
      particle close to the bounding surface.
The solution to these problems is essential to understand the behavior 
of magnetic or nonmagnetic particles of colloidal size inherently 
present in biological liquid-crystalline tissues like cellular membranes.
These problems also arise in ferroliquid crystals -- the suspensions 
of single-domain ferroparticles in liquid crystals \cite{brochard.f:1970.a}. 

The answer to the first question is known for spherical colloidal 
particles with homeotropic anchoring of the director at the particle surface \cite{lubensky.tc:1998.a,poulin.p:1997.a,poulin.p:1998.a,kuksenok.ov:1996.a,shiyanovskii.sv:1998.a,mondain-monval.o:1999.a,ruhwandl.rw:1997.b,stark.h:1999,stark.h:2001.a}. 
Isolated particles provide a spherical confining geometry 
for the liquid crystal. Sufficiently strong homeotropic anchoring induces 
a hedgehog defect with topological charge $+1$. The total topological 
charge of the whole system is zero, and an additional defect must be created 
to compensate the radial hedgehog. Two types of defect are possible:
a hyperbolic hedgehog with a topological charge $-1$, 
called a dipolar or satellite defect; 
or a $-\frac{1}{2}$ strength disclination ring that encircles 
the spherical particle, called a quadrupolar or Saturn-ring defect. 
The dipolar (satellite) defect is a point defect, while the quadrupolar 
(Saturn ring) is a line defect.
Theoretical and numerical work based on elastic theory 
\cite{ruhwandl.rw:1997.b,stark.h:1999}, as well as 
computer simulation \cite{billeter.jl:2000.a,andrienko.d:2001.b} ,
suggests that the dipole configuration is stable for micron-sized 
droplets. It is the one usually realized experimentally. 
The Saturn-ring defect appears if the droplet size is reduced 
or an external field is applied \cite{gu.y:2001.a,loudet.jc:2001.a}.

The same topological arguments are applicable for a non-spherical colloidal
particle with homeotropic anchoring of the director at the particle surface. 
For an elongated particle with length $L$ and transverse 
size $D << L$, and both $L$ and $D$ much greater than the dimensions of the 
molecules of the liquid crystal, one can have a disclination 
line of strength $-1$, a pair of disclination lines of 
strength $-\frac{1}{2}$, 
as well as the `escaped radial' structure, in which the 
director bends over to become perpendicular to the particle surface
\cite{burylov.sv:1994.a}.

However, from the energetic point of view, the situation is different
from the case of the spherical particle. 
For the elongated particle, both defects are disclination {\em lines}. 
The elastic energy per unit length associated with a disclination 
 of strength $m$ is $\pi Km^2 \ln(R/r_0)$, where $R$ is the size of 
the sample and $r_0$ 
is a lower cutoff radius (the core size) \cite{stephen.mj:1974.a}. 
This means that the free energy of a pair of $-\frac{1}{2}$ disclinations 
is always smaller than that of a single $-1$ disclination. 
Therefore, one can expect that the pair
of $-\frac{1}{2}$ disclinations will always be a stable configuration. 
In principle, the $-1$ defect can still form a metastable state.

The answer to the second question is not known even in the framework of 
phenomenological (continuum) theory \cite{burylov.sv:1994.a}. 
The results of the theory only indicate that, depending 
on the type and strength of anchoring, the equilibrium position of the 
particle may be either parallel or perpendicular to the liquid crystal 
director. The parameter governing the situation is the ratio of the 
particle radius to the extrapolation length of the nematic liquid crystal. 

Obtaining an analytical expression for the torque (or elastic free energy) 
at arbitrary tilt angle $\theta$ seems to be hardly possible
due to the loss of symmetry of the director distribution 
and the presence of defects. 
Qualitative analysis shows that the proper argument for the free energy 
should be $({\bm n}_0 \cdot {\bm n})^2 = \cos^2\theta$ since the problem 
is bilinear in both the unperturbed director orientation ${\bm n}_0$ 
and the unit vector along the symmetry axis of the rod, ${\bm n}$. 
A simple form of the free energy angular dependence has been proposed 
in Ref.~\cite{burylov.sv:1994.a}
\begin{equation}
{\cal F}(\cos^2\theta) = {\cal F}_{\perp} + 
({\cal F}_{\parallel} - {\cal F}_{\perp}) \cos^2\theta,
\label{free_en_rod}
\end{equation}
which gives a $\sin 2\theta$ dependence for the torque and predicts
that the director response has a maximum at $\theta = \pi / 4$ and is absent 
at $\theta = 0, \pi / 2$.  
However, it is clear that eqn~(\ref{free_en_rod}) is oversimplified. 
The defect structure 
changes while the particle rotates. The nematic ordering evolves
in a complicated way that can hardly be approximated with a $\sin 2\theta$
dependence of the torque. At the same time, this dependence is vital
for a macroscopic description of the system which treats coupling of 
colloidal particles with the nematic host via an effective potential. 

Recently, the effect of confinement on the orientation of an anisotropic
colloidal particle has been predicted \cite{roth.r:2001.a}.
It has been shown that there is an `entropic' 
torque on a hard rod-like particle dissolved in a solution
of hard spheres when the rod is positioned close to the hard wall.
The torque appears because of the density modulation of spheres near 
the wall and depletion forces between the wall and the hard rod.
This torque might play an important role in the `key-lock' principle
in biological systems and provide an understanding of how a non-spherical 
`key' macromolecule can adjust its position and orientation near the 
`lock' macromolecule. 
We, therefore, expect the torque on the particle close to the wall 
to be different from the bulk-induced torque because of the additional, 
entropic contribution.

In this paper, we present the results of molecular dynamics (MD) and 
Monte Carlo (MC) simulations of the topological defects that appear in 
the nematic mesophase around an elongated colloidal particle. 
Using the MD technique, we also study the force and the torque on
the particle suspended in the bulk of the nematic mesophase and the 
modification of this torque when the particle is close to the cell substrate.

The paper is organized as follows. In section~\ref{model} we present
the computational details and molecular models we use to simulate the 
liquid crystal mesophase and the interaction of the molecules with the
particle surface and the cell substrates. Section \ref{results}
contains the results of the MD and MC simulations: 
density, director, order parameter maps, and order tensor profiles of 
the defects. Here we also present the results for the torque on the 
particle in the cell bulk and near the wall.
Concluding remarks and comparison of the techniques are given in 
section~\ref{conclusions}.

\section{Molecular model and simulation methods}
\label{model}
\begin{figure}
\begin{center}
\includegraphics[width=4cm, angle=0]{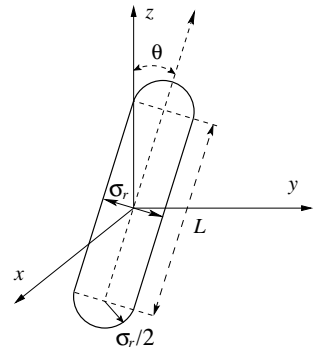}
\end{center}
\caption[Studied geometry]{
\label{fig:1} % fig1.ps
Studied geometry: a spherocylinder of length $L$ and diameter $\sigma_r$ 
is immersed in a liquid crystal host, which is modeled either as a solution 
of Gay-Berne particles (molecular dynamics) or `spins' fixed on a cubic 
lattice (Monte Carlo simulations). 
The symmetry axis of the spherocylinder is tilted with respect 
to the $z$ axis. To study the defect structure (sec.~\ref{sec:defects}) 
we use a rod of infinite length positioned along the $y$ axis, normal to 
the director. 
}
\end{figure}

Figure \ref{fig:1} shows the geometry of a single particle in the cell. 
The rod shape is chosen to be a spherocylinder, i.e. a cylinder of 
length $L$ and diameter $\sigma_r$ which has spherical caps of 
diameter $\sigma_r$. The particle is tilted in the $zy$ plane.
The orientation of the particle is specified by the angle $\theta$ between 
the $z$ axis and the symmetry axis of the cylinder.  

\subsection{Molecular dynamics}
Molecular dynamics simulations were carried out using the soft repulsive 
potential, describing (approximately) ellipsoidal molecules
\begin{equation}
\label{eqn:pot}
v_{ij} = \left\{\begin{array}{lr}
4 \varepsilon_0\left(\varrho_{ij}^{-12}-\varrho_{ij}^{-6} \right) +
 \varepsilon_0, & \varrho_{ij}^6 < 2 \\
0 & \varrho_{ij}^6 > 2
\end{array} \right. \:.
\end{equation}
Here $\varrho_{ij} = (r_{ij}-\sigma_{ij}+\sigma_0)/\sigma_0$; $r_{ij}$ is the
center-center separation, $\sigma_0$ a size parameter, $\varepsilon_0$ an 
energy parameter (both taken to be unity) and the orientation-dependent 
diameter $\sigma_{ij}$ is defined by
\begin{eqnarray}
\nonumber
1- \frac{\sigma_0^2}{\sigma_{ij}^2} =
\frac{\chi}{2}\left[
\frac{(\hat{\bm r}_{ij}\cdot {\bm u}_i
     + \hat{\bm r}_{ij}\cdot {\bm u}_j)^2}
     {1 + \chi({\bm u}_i\cdot {\bm u}_j)} + 
%\right. \right. \nonumber \\ && \left. \left.
\frac{(\hat{\bm r}_{ij}\cdot {\bm u}_i
     - \hat{\bm r}_{ij}\cdot {\bm u}_j)^2}
     {1 - \chi({\bm u}_i\cdot {\bm u}_j)}
\right],
\end{eqnarray}
where $\chi=(\kappa^2-1)/(\kappa^2+1)$, $\kappa$ being the elongation.  In this
work we used $\kappa=3$ throughout. The orientation dependence is written in
terms of the direction of the center-center vector $\hat{\bm r}_{ij} = {\bm
r}_{ij}/r_{ij}$ and the unit vectors ${\bm u}_i$, ${\bm u}_j$ which
specify the molecular symmetry axes.  The potential  (\ref{eqn:pot}) 
may be thought of as a variant of the standard Gay-Berne potential 
\cite{berne.bj:1972.a,gay.jg:1981.a} with exponents $\mu = 0, \nu = 0$. 
 
The systems consisted of $N = 64,000$ particles. 
A reduced temperature $k_{\rm B}T/\epsilon_0=1$ 
was used throughout (for this model, the phase behavior is not 
sensitively dependent on temperature, as there are no attractive forces).  
The system size was chosen so that the number density of
the liquid crystal far from the rod was $\rho\sigma_0^3 \approx 0.34$. 
For this system, in the reduced units defined by $\sigma_0$,
$\epsilon_0$ and $m$, a timestep $\delta t=0.004$ was found suitable.
The molecular moment of inertia was fixed as $I=2.5m\sigma_0^2$.
Periodic boundary conditions as well as slab geometry with walls
confining the system in the $z$ direction were considered.  

The interaction of molecule $i$ with the rod
was given by a shifted Lennard-Jones repulsion potential having exactly 
the same form as eqn~(\ref{eqn:pot}), but with $\rho_{ij}$
replaced by 
$\rho_{i} = (|{\bm r}_{i}-{\bm r}_s|-\sigma_{r}/2+\sigma_0/2)/\sigma_0$. 
Here ${\bm r}_s = \gamma_i {\bm n}$, where 
$\gamma_i = {\rm sign}({\bm n} \cdot {\bm r}_i + z_r \cos \theta)
\min\left(L/2, |{\bm n} \cdot {\bm r}_i + z_r \cos \theta|\right)$, 
${\bm n} = (0, \sin\theta, \cos\theta)$ 
is a unit vector along the symmetry axis of the rod, 
$z_r$ is the distance from the center of the rod to the center of 
the coordinates.
In slab geometry the interaction with the walls was given by the same 
formula, eqn~(\ref{eqn:pot}), replacing particle $j$ by the wall $w$,
setting $\rho_{iw} = (|z_{iw}| - \sigma_{iw}/2 + \sigma_{0}/2)/\sigma_{0}$ 
and $\sigma_{iw}^2 = \kappa_w^2 +(1-\kappa_w^2)(1-e_{iz}^2)$. 
$\kappa_w$ represents an effective particle elongation as seen by the wall.
We used $\kappa_w = 1$ which gives strong homeotropic orientation of 
the molecules at the wall \cite{andrienko.d:2001.b}.

The radius and length of the rod were steadily increased from zero to
the desired value during $10^3$ steps. Then the system was equilibrated 
for $10^6$ steps. During equilibration we scaled the velocities of 
the molecules to achieve $k_{\rm B}T/\epsilon_0 = 1$. 

The production run for every tilt angle of the rod was $10^6$ steps.
The force ${\bm F}$ and the torque ${\bm M}$ on the rod were calculated 
using the repulsive force ${\bm f}_i$ from the rod on the particle $i$
\begin{eqnarray}
\nonumber
{\bm F} &=& - \sum_{i=1}^N  {\bm f}_{i}, \\
\label{torque}
{\bm M} &=& - \sum_{i=1}^N \gamma_i \left[ {\bm f}_{i} \times {\bm n} \right].
\end{eqnarray}

\subsection{Monte Carlo simulation}
Monte Carlo simulations were based on the Lebwohl-Lasher 
(LL) lattice model~\cite{lebwohl.pa:1972.a}. Within this model uniaxial 
nematic molecules (or, alternatively, close-packed molecular
clusters~\cite{berggren.e:1994.b}) are represented by `spins' fixed on a
cubic lattice of spacing $a$. The spins are allowed to rotate freely, 
which reproduces the orientational behavior of the liquid crystal 
sufficiently well. 

To define the topology of the colloidal particle, a `jagged' cylinder
of diameter $\sigma_r$ was carved from the cubic lattice, with its long
axis fixed along the $y$ axis of the coordinate system.
The orientations of spins representing the particle were kept fixed during
the simulation and were chosen in agreement with the desired boundary
conditions at the particle surface, in our case homeotropic. At outer
boundaries of the simulation box periodic boundary conditions were assumed.
The total interaction energy for our model system consisting of
nematic spins was calculated as
\begin{eqnarray}
 U = -\sum_{\langle i<j\rangle}\epsilon_{ij}\,P_2(\cos\beta_{ij}), 
\label{energy_ll}
\end{eqnarray}
with $P_2(x)=\frac{1}{2}(3x^2-1)$ and $\cos\beta_{ij}={\bm u}_i\cdot{\bm
u}_j$. Here ${\bm u}_i$ denotes the unit vector giving the orientation of
the spin located at the $i$th lattice site. 
The sum in eqn~(\ref{energy_ll}) is taken over nearest neighbors only. 
The $\epsilon_{ij}$ constants represent the interaction strengths and 
are denoted by $\epsilon$ and $\epsilon_p$ for nematic-nematic and 
nematic-solid particle interactions respectively.

The simulation box size was set to $30\,a\times 30\,a\times 30\,a$, 
which for the chosen cylinder diameter ($\sigma_r=10\,a$) amounts to
24600 nematic spins and to 840 spins representing the surface of the solid
particle. Our simulations started from a configuration with a random orientation
of nematic spins. The final results did not depend on the choice of the 
starting configuration.

The standard Metropolis scheme~\cite{metropolis.n:1949.a} was then employed to 
update nematic spin orientations~\cite{fabbri.u:1986.a,barker.ja:1969.a},
maintaining a rejection ratio close to 0.5. The system was equilibrated
during $\approx 6\times 10^4$ MC cycles. After this $6.6\times 10^4$
successive spin configurations were accumulated and used as input for
the calculation of order tensor.

In the simulation, temperature was set to $T^*=k_BT/\epsilon=1$, which
ensures the existence of the nematic phase (note that for a bulk sample
the LL model exhibits a nematic-isotropic transition at 
$T^*=1.1232$~\cite{fabbri.u:1986.a}). The strengths of nematic-nematic
and nematic-solid particle interactions were set equal,
$\epsilon_p=\epsilon$, which corresponds to the strong anchoring
regime with the extrapolation length of the order of a few lattice
spacings $a$~\cite{priezjev.n:2000.a}.

\subsection{Order tensor}
For both techniques, the local order tensor $\bm{Q}(\bm{r})$ was calculated 
\begin{equation}
\label{order_tensor}
Q_{\alpha \beta} = \frac{1}{n} \sum_{k=1}^{n}
\left\{ \frac{3}{2}\left<u_{k \alpha}u_{k \beta}\right> - 
\frac{1}{2} \delta_{\alpha \beta} \right\},
\end{equation} 
where there are $n$ molecules present in each bin, $\delta_{\alpha \beta}$ 
is the Kronecker delta, $\left< \cdots \right>$ denotes an ensemble average,
$\alpha, \beta = x, y, z$. Note that in the MC case we have $n=1$ (bins correspond
to lattice points) and that averaging is performed over MC cycles only.
Diagonalizing the $Q_{\alpha\beta}$ tensor, for each bin, gives three eigenvalues,
$Q_1$, $Q_2$, and $Q_3$, plus the three corresponding eigenvectors. The eigenvalue
with the largest absolute value defines the order parameter $S$ for each bin. The
biaxiality $P$ is then calculated as the absolute value of the difference between
the remaining two eigenvalues of the order tensor $\bm Q$.

\section{Simulation results and discussion}
\label{results}
\subsection{Defect structure}
\label{sec:defects}
We started our simulations with a rod
of infinite length, $L = \infty$, positioned along the $y$ axis, 
normal to the director. In this case the director rotates in the $xz$ 
plane and we effectively have a two-dimensional situation. 

\subsubsection{MC results}
\begin{figure}
\includegraphics[width=6cm]{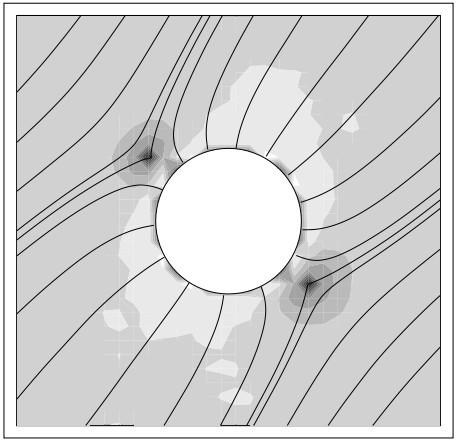}
\caption{
MC simulation results: cross section of the director field ${\bm n}(x,z)$.
The shading represents the value of the order parameter, $S$.
A pair of $-\frac{1}{2}$ defects has formed on the diagonal.
In the defect core molecules are (on the average) aligned in the $xz$ plane; 
ordering is uniaxial with $S<0$ and the corresponding eigenvector, 
$\bm n$, is directed out-of-plane (along the long axis of the particle).}
\label{fig:mc_order}
\end{figure}

Fig.~\ref{fig:mc_order} shows the director field and the order parameter
map, in the plane perpendicular to the long axis of the colloidal particle. 

\begin{figure}
\includegraphics[width=8cm]{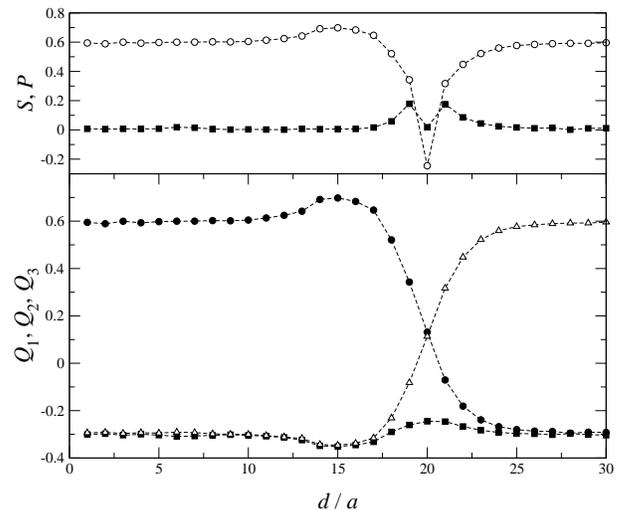}
\caption{MC simulation results: order tensor components 
$Q_{1}$ (triangles), $Q_{2}$ (circles), and $Q_{3}$ (squares) 
plotted across the defect. In the upper panel we plot the order parameter
$S$ (open circles) and biaxiality $P$ (squares).
The left-right asymmetry with respect
to the defect core is due to the presence of the colloidal particle.}
\label{fig:mc_tensor}
\end{figure}

As concluded from topological considerations, either a $-1$ strength
disclination line or a pair of $-\frac{1}{2}$ lines can form in the
neighborhood of the particle. The $-1$ line, however, does not seem to
be stable and splits into a pair of $-\frac{1}{2}$ lines during the
MC evolution, even if it is taken as initial configuration in the
simulation run. This behavior agrees with simple estimates
of defect line free energies. Moreover, a stable `escaped radial'
structure was also not observed in our simulations.
In MC simulations the pair of defect lines always forms close to one of
the simulation box diagonals although the cross section of the colloidal
particle is axially symmetric (ignoring its jagged shape); see
the director field shown in Fig.~\ref{fig:mc_order}. This symmetry
breaking may be attributed to two effects of different origin. The first
one (and, according to our tests, the more important one) is the repulsion
between defects maximizing the defect-to-defect distance (recall the periodic
boundary conditions), while the second one is a finite-size effect
originating from collective fluctuations, resulting in a tendency to
align the nematic along the simulation box diagonal~\cite{braun.fn:2001.a}.
We believe, however, that these phenomena, as well as the presence of
the colloidal particle, do not considerably affect any of the qualitative
features characterizing the disclination line inner structure. Moreover,
the presence of the colloidal particle is reflected only in an enhancement
of the degree of nematic order in the immediate surroundings of the particle.
The inner structure of a defect line is further characterized by variations
in order tensor components, $Q_{1}$, $Q_{2}$, and $Q_{3}$, obtained after
diagonalization of the order tensor $\bm Q$, eqn~(\ref{order_tensor}).
This is the most convenient way of describing the structure of the defect,
because of the possible biaxiality and negative values of the uniaxial order
parameter in the core region.

Fig.~\ref{fig:mc_tensor} shows the $Q_{1}$, $Q_{2}$, and $Q_{3}$ profiles
plotted along the $z$ axis through the left of the two disclinations shown in
Fig.~\ref{fig:mc_order}. In Fig.~\ref{fig:mc_tensor} the disclination is
located at $d=20a$. Note that the left-right assymetry of the profiles with
respect to the defect position is caused solely by the presence of the
colloidal particle. As shown by Fig.~\ref{fig:mc_tensor}, the $Q_1$ component
changes from its positive bulk value ($\approx 0.6$), coinciding with the value
of the order parameter $S$) to some negative value
($\approx -0.3$) after passing through the disclination. At the same time,
the $Q_2$ component increases from a negative value ($\approx -0.3$) 
to a large positive value ($\approx 0.6$), which roughly equals twice the 
absolute value of the negative one. This behavior is attributed to the director
rotation by approximately $\pi/2$ when we cross the defect along the $z$ axis;
see Fig.~\ref{fig:mc_order}. On the other hand, the value of the $Q_3$
component does not change too much, indicating that the variation in the
nematic ordering mostly occurs in the $xz$ plane, perpendicular to the
symmetry axis of the particle.
Alternatively, $Q_{1}$, $Q_{2}$, and $Q_{3}$ profiles can be interpreted also
in terms of order parameters $S$ and $P$ (see Fig.~\ref{fig:mc_tensor}, 
upper panel). When the defect line is approached,
the uniaxial order parameter $S$ decreases from its temperature-defined bulk
value and drops even below zero in the defect center. Note that there
the nematic director, i.e., the eigenvector corresponding to the negative
eigenvalue, is directed along the long axis of the solid particle. On the
other hand, the biaxiality -- close to zero far enough from the defect --
increases when the defect line is approached, reaches a maximum and, finally,
in the very center of the defect, again drops to a value which is close to 
zero.
The characteristic length scales for these variations are of the order of
a few ($\approx 5$) lattice spacings $a$ and agree with the estimates for
the correlation length in the nematic phase~\cite{fabbri.u:1986.a}.

Qualitatively, molecular ordering close to a disclination line can
be summarized as follows. In the very center of the defect molecular
ordering is uniaxial with $S<0$ and $P\to 0$. Far enough from the defect
line the nematic liquid crystal is uniaxial again, however, with $S>0$
and $P=0$, as expected in a homogeneous or in a weakly distorted bulk sample.
In the intermediate ring-like region, nematic ordering is biaxial with $P\neq 0$.
These conclusions agree also with results from alignment tensor-based
phenomenological analyses of topological defects both
of half-integer~\cite{schopohl.n:1987.a} and integer
strength~\cite{sonnet.a:1995.a}.

\subsubsection{MD results}

\begin{figure}
\begin{center}
\includegraphics[width=6cm]{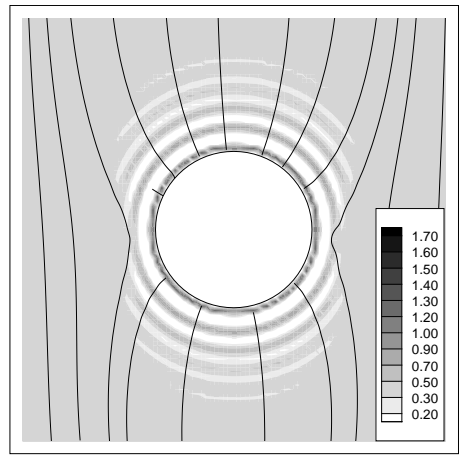}
\caption[]
{MD simulation results: director streamlines of the $xz$ cross section of the 
director field. Rod diameter $\sigma_r = 20\sigma_0$, rod length $L = \infty$.
The shading represents the value of the density. 
The director far from the particle is constrained along the $z$ axis.
A pair of $-\frac{1}{2}$ line defects forms parallel to the 
particle axis, perpendicular to the director far from the particle. 
}
\label{fig:md_density}
\end{center}
\end{figure}

\begin{figure}
\begin{center}
\includegraphics[width=8cm]{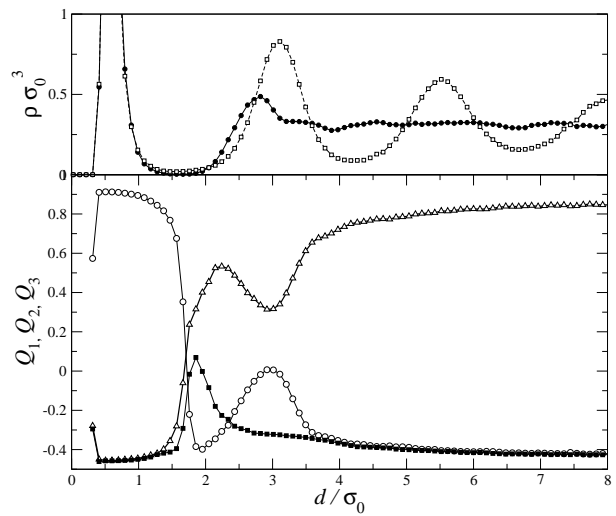}
\caption[]
{MD simulation results: order tensor components 
$Q_{1}$ (triangles), $Q_{2}$ (circles), and $Q_{3}$ (squares)
across the defect. 
In the upper panel we plot the density profile across the defect (circles) 
and avoiding the defect (open squares).
The density modulation near the particle affects
the order parameter variation in the core region. 
}
\label{fig:md_tensor}
\end{center}
\end{figure}

As has already been mentioned, the configuration with two $-\frac{1}{2}$ 
disclination lines is more energetically favorable than 
a single $-1$ strength disclination.
We noticed this performing the molecular dynamics simulation: for all studied
diameters of the rod ($\sigma_r/\sigma_0 = 5-20$) the $-1$ strength 
disclination appears immediately after expanding the colloid particle 
in the nematic state. However, during the equilibration, it splits into 
two $-\frac{1}{2}$ disclination lines which then move towards 
the equatorial plane. The evolution dynamics 
is quite slow, one needs about $10^6$ steps for the $-\frac{1}{2}$ 
disclinations to reach the equator.
We were not able to observe `escaped radial' configuration, probably due to 
the small size of the colloidal particle. 

A typical director map together with the density map is shown 
in Fig.~\ref{fig:md_density}. 
The bulk density, $\rho_{\rm b} \sigma_0^3 \approx 0.34$, 
is slightly different from the number density, $\rho \sigma_0^3 = N/V = 0.33$,
because of the volume taken by the rod. 
In the direction of the disclinations the density modulation, 
typical for a nematic-wall interface, vanishes due to 
partial melting of the liquid crystal in the disclination core region. 
This melting damps the influence of the droplet surface on the 
interface region. 

Two $-\frac{1}{2}$ disclinations are located very close to 
the droplet surface and the director distortion vanishes very quickly 
in the liquid crystal bulk. The core region extends over a few molecular 
lengths. In MD simulations, the pair of defect lines forms perpendicular
to the director. Since we use the director constraint algorithm
\cite{germano.preprint}, the director far from the particle is aligned along
the $z$ axis, contrary to the situation with MC simulation results,
where the director is along the box diagonal. The director constraint
damps the effects of the defect repulsion and effective fluctuation of the 
director.

To emphasize the complex structure of the defect 
core we plot the order tensor components $Q_{1}$, $Q_{2}$, $Q_{3}$, after 
diagonalizing the local order tensor, eqn~(\ref{order_tensor}),
in Fig.~\ref{fig:md_tensor}. Qualitatively, the order tensor has the same 
look as in MC simulations: the nematic phase is uniaxial far from the core 
and biaxial in the core region, with variation of the biaxiality 
across the core. However, MD results predict a more complicated
structure of the core region, attributed to the liquid crystal density 
oscillation near the particle surface (see Fig.~\ref{fig:md_density} and
Fig.~\ref{fig:md_tensor}, upper panel). 
The variation of the order tensor components is given by the superposition of 
the nematic order variation due to the density modulation and intrinsic 
variation due to the presence of the defect.

\subsection{Torque on the particle}
To measure the torque on the rod of finite length, we performed 
MD simulations in a box with 
periodic boundary conditions, applying a global constraint for 
the director along the $z$ axis \cite{germano.preprint}. 
An independent measurement was performed in slab geometry.
In the slab geometry, the director orientation far from the rod 
was fixed by the confining walls. 
The walls provided strong homeotropic (along the $z$ axis) anchoring of 
the director. 
The rod was fixed either in the center of the simulation box or 
at some distance $z$ from the bottom wall.
$M_x > 0$ ($< 0$) corresponds to a torque which tends to align the rod
perpendicular (parallel) to the director far from the particle, 
${\bm n}_0 = {\bm e}_z$.

\begin{figure}
\begin{center}
\includegraphics[width=8cm]{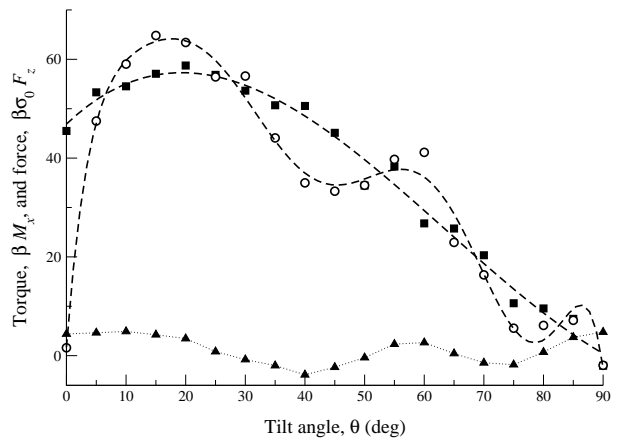}
\caption[Torque on the rod vs. rod tilt angle]{ 
Torque on the rod vs. rod tilt angle. 
$M_x > 0$ ($< 0$) corresponds to a torque which tends to align the rod
perpendicular (parallel) to the director far from the particle, 
${\bm n}_0 = {\bm e}_z$.
Rod diameter $\sigma_r = 5\sigma_0$, rod length $L = 10\sigma_0$.
Squares - slab geometry, with the rod in the middle of the cell.
Circles - slab geometry, with the center of the particle located 
at the distance $z = 15\sigma_0$ from the bottom wall. 
Triangles - the depletion force on the particle, when it is near the wall.
As a guide, the dashed lines correspond to a polynomial fit.
See also Fig.~\ref{fig:md_dir} for explanations.
}
\label{fig:md_torque}
\end{center}
\end{figure}

The torque on the rod, calculated using eqn~(\ref{torque}), 
is presented in Fig.~\ref{fig:md_torque}.
Results presented in Fig.~\ref{fig:md_torque} indicate that the dependence of 
the torque on the rod tilt angle is far from the $\sin 2\theta$, 
proposed in \cite{burylov.sv:1994.a}.
Moreover, the torque is not equal to zero for $\theta = 0^0$,
i.e. there is some symmetry breaking and 
the orientation of the rod along the director is not even metastable. 

\begin{figure}
\begin{center}
\includegraphics[width=8cm]{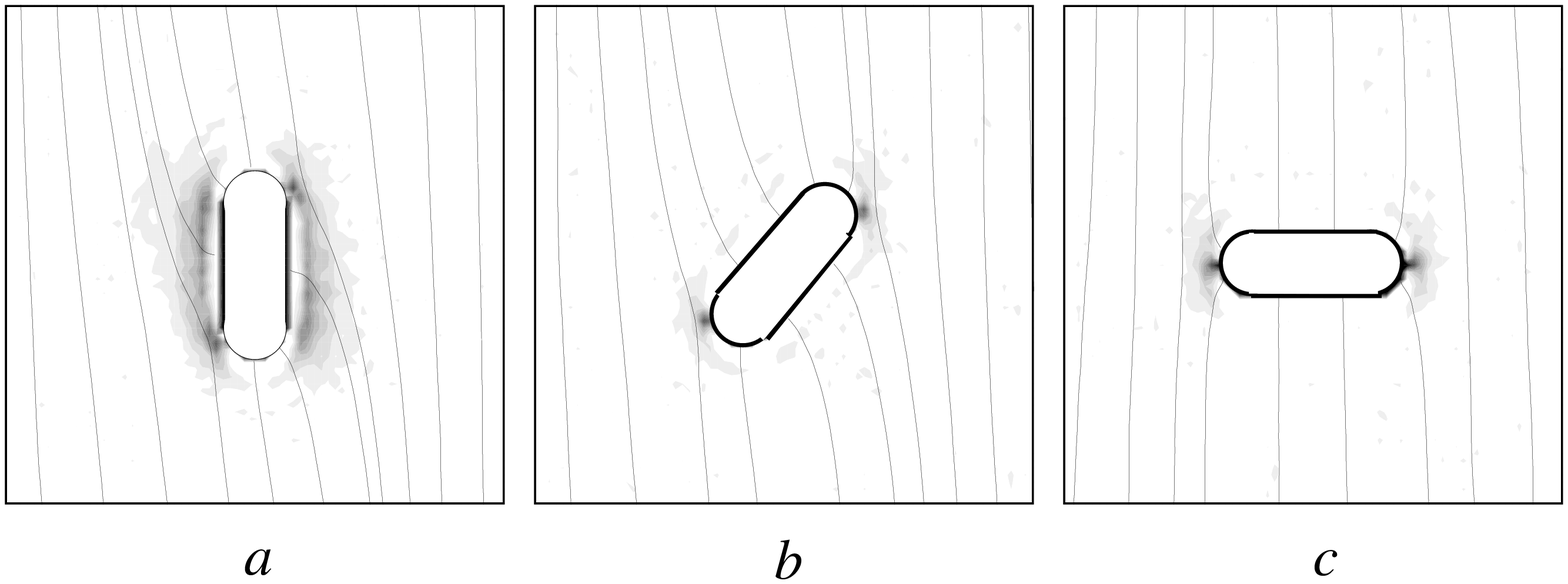}
\caption[Director streamlines and order parameter maps 
         for different tilt angles]
{ 
Director streamlines and order parameter maps for different tilt angles: 
a) $\theta = 0^0$; b) $\theta = 45^0$; c) $\theta = 90^0$.
A side view along the $x$ axis is shown (the rod is tilted in the $yz$ plane). 
Rod diameter $\sigma_r = 5\sigma_0$, rod length $L = 10\sigma_0$.
}
\label{fig:md_dir}
\end{center}
\end{figure}

For better understanding, a slice in the $yz$ plane is shown 
in Fig.~\ref{fig:md_dir}, for different tilt angles of the rod. 
Fig.~\ref{fig:md_dir}a shows that the director distribution around the rod
is not axially symmetric. This is the reason for the 
non-zero torque for $\theta = 0^0$. Strong director variations near the
rod are responsible for the large value of the torque. As the 
rod rotates, the director field becomes less and less frustrated,
Fig.~\ref{fig:md_dir}b, and finally we have a stable orientation of the 
rod perpendicular to the director, Fig.~\ref{fig:md_dir}c. 

In principle, the configuration with axial symmetry is also possible,
when the rod is along the $z$ axis. However, we were not able to observe it
in our simulations even when disordered, isotropic configurations  
containing the colloidal particle were compressed to the ordered, nematic
state. This method, in principle, gives the lowest free-energy configurations
in an unbiased way.

In Fig.~\ref{fig:md_torque} we also present the torque and the force on 
the particle near the wall. Besides the torque on the particle
because of the average molecular orientation of the liquid crystal host, 
there is an {\em effective} interaction between the particle and the wall. 
The presence of this depletion-like interaction can be understood as a 
result of interactions between the particle and the liquid crystal 
molecules which themselves interact with the wall.
In other words, if the volume close to both the wall and the particle
overlap, then the host liquid gains accessible volume and can increase 
its entropy. To show that the depletion force is responsible for this 
`entropic' contribution to the torque, we plot the depletion force
$\beta \sigma_0 F_z (\theta)$ in Fig.~\ref{fig:md_torque}. The correlation
between the change in the depletion force and the change in the torque 
is evident.
As the rod moves closer to the wall the value of the `entropic' torque
increases and can even affect the equilibrium position and 
orientation of the particle.

For the parameters used in our simulation, the contribution of the `entropic'
torque only modifies the dependence of the total torque on the particle 
tilt angle. However, this contribution can dominate, for example, 
when the system is close to the nematic-isotropic transition.
Then the colloidal particle might have a tilted orientation when approaching 
the wall. Further work on these aspects is in progress.

\section{Conclusions}
\label{conclusions}
We used molecular dynamics and Monte Carlo techniques to study a small 
elongated colloidal particle suspended in a nematic liquid crystal. 
Homeotropic boundary conditions and strong anchoring  create 
a hedgehog defect on the particle surface. We have studied the defect 
structure around the particle that cancels this hedgehog defect. 

Our simulation results show that, in the case of a very long particle, 
with transverse size much less than its length, a configuration
with two $-\frac{1}{2}$ defects is stable. An initial configuration
with one $-1$ strength disclination evolves spontaneously into two
 $-\frac{1}{2}$ disclination lines. 
For a particle with transverse size of the order of its length,
the $-\frac{1}{2}$ ring defect encircling the particle was found to be 
stable. The orientation of the ring changes as the particle tilts
with respect to the director.
Using order tensor and density maps we are able to 
resolve the structure of the core of the defect: the nematic phase is
strongly biaxial near the defect core. 

Comparing the structure of the disclination core obtained using 
molecular dynamics and Monte Carlo simulation techniques 
we are able to study influence of the density modulation of the mesophase in 
the core region.

We have also studied the torque on a particle tilted with respect to
the director at large distances from a wall, and modification of this torque
when the particle is close to the wall. Our results show, that the 
dependence of the torque on the tilt angle is complex,
and does not vanish even when the particle is along the director. 
Analysis of the director distribution around the particle shows that 
this is due to the broken axial symmetry of the director distribution 
around the particle.
In addition, when the particle is close to the wall, the torque is 
modified because of the depletion-like interaction of the particle
with the wall.

\begin{acknowledgments}
This research was supported by EPSRC grants GR/L89990, GR/M16023,
through INTAS grant 99-00312, and by the Ministry of Science and
Technology of Slovenia (Grant No. J1-7470).
MD simulations used the GBMEGA program of the `Complex Fluids
Consortium' with computer time allocated at the Edinburgh Parallel 
Computer Center and the CSAR facility.
D.A. acknowledges the support of the 
Overseas Research Students Grant; M.P.A. is grateful to the Alexander 
von Humboldt foundation, the British Council ARC program, and the 
Leverhulme Trust. G.S. thanks P. Pasini, C. Zannoni, and C. Chiccoli
for the training in MC simulation techniques.
\end{acknowledgments}

\bibliography{journals,main,extra}

\begin{thebibliography}{36}
\expandafter\ifx\csname natexlab\endcsname\relax\def\natexlab#1{#1}\fi
\expandafter\ifx\csname bibnamefont\endcsname\relax
  \def\bibnamefont#1{#1}\fi
\expandafter\ifx\csname bibfnamefont\endcsname\relax
  \def\bibfnamefont#1{#1}\fi
\expandafter\ifx\csname citenamefont\endcsname\relax
  \def\citenamefont#1{#1}\fi
\expandafter\ifx\csname url\endcsname\relax
  \def\url#1{\texttt{#1}}\fi
\expandafter\ifx\csname urlprefix\endcsname\relax\def\urlprefix{URL }\fi
\providecommand{\bibinfo}[2]{#2}
\providecommand{\eprint}[2][]{\url{#2}}

\bibitem[{\citenamefont{Lubensky et~al.}(1998)\citenamefont{Lubensky, Pettey,
  Currier, and Stark}}]{lubensky.tc:1998.a}
\bibinfo{author}{\bibfnamefont{T.~C.} \bibnamefont{Lubensky}},
  \bibinfo{author}{\bibfnamefont{D.}~\bibnamefont{Pettey}},
  \bibinfo{author}{\bibfnamefont{N.}~\bibnamefont{Currier}}, \bibnamefont{and}
  \bibinfo{author}{\bibfnamefont{H.}~\bibnamefont{Stark}},
  \bibinfo{journal}{Phys.\ Rev.\ E} \textbf{\bibinfo{volume}{57}},
  \bibinfo{pages}{610} (\bibinfo{year}{1998}).

\bibitem[{\citenamefont{Stark}(1999)}]{stark.h:1999}
\bibinfo{author}{\bibfnamefont{H.}~\bibnamefont{Stark}},
  \bibinfo{journal}{Euro.\ Phys.\ J. B} \textbf{\bibinfo{volume}{10}},
  \bibinfo{pages}{311} (\bibinfo{year}{1999}).

\bibitem[{\citenamefont{Lev and Tomchuk}(1999)}]{lev.bi:1999.a}
\bibinfo{author}{\bibfnamefont{B.~I.} \bibnamefont{Lev}} \bibnamefont{and}
  \bibinfo{author}{\bibfnamefont{P.~M.} \bibnamefont{Tomchuk}},
  \bibinfo{journal}{Phys.\ Rev.\ E} \textbf{\bibinfo{volume}{59}},
  \bibinfo{pages}{591} (\bibinfo{year}{1999}).

\bibitem[{\citenamefont{Bor{\v s}tnik et~al.}(1999)\citenamefont{Bor{\v s}tnik,
  Stark, and {\v Z}umer}}]{borstnik.a:1999.a}
\bibinfo{author}{\bibfnamefont{A.}~\bibnamefont{Bor{\v s}tnik}},
  \bibinfo{author}{\bibfnamefont{H.}~\bibnamefont{Stark}}, \bibnamefont{and}
  \bibinfo{author}{\bibfnamefont{S.}~\bibnamefont{{\v Z}umer}},
  \bibinfo{journal}{Phys.\ Rev.\ E} \textbf{\bibinfo{volume}{60}},
  \bibinfo{pages}{4210} (\bibinfo{year}{1999}).

\bibitem[{\citenamefont{Bor{\v s}tnik et~al.}(2000)\citenamefont{Bor{\v s}tnik,
  Stark, and {\v Z}umer}}]{borstnik.a:2000.a}
\bibinfo{author}{\bibfnamefont{A.}~\bibnamefont{Bor{\v s}tnik}},
  \bibinfo{author}{\bibfnamefont{H.}~\bibnamefont{Stark}}, \bibnamefont{and}
  \bibinfo{author}{\bibfnamefont{S.}~\bibnamefont{{\v Z}umer}},
  \bibinfo{journal}{Phys.\ Rev.\ E} \textbf{\bibinfo{volume}{61}},
  \bibinfo{pages}{2831} (\bibinfo{year}{2000}).

\bibitem[{\citenamefont{Poulin et~al.}(1997)\citenamefont{Poulin, Stark,
  Lubensky, and Weitz}}]{poulin.p:1997.a}
\bibinfo{author}{\bibfnamefont{P.}~\bibnamefont{Poulin}},
  \bibinfo{author}{\bibfnamefont{H.}~\bibnamefont{Stark}},
  \bibinfo{author}{\bibfnamefont{T.~C.} \bibnamefont{Lubensky}},
  \bibnamefont{and} \bibinfo{author}{\bibfnamefont{D.~A.} \bibnamefont{Weitz}},
  \bibinfo{journal}{Science} \textbf{\bibinfo{volume}{275}},
  \bibinfo{pages}{1770} (\bibinfo{year}{1997}).

\bibitem[{\citenamefont{Poulin and Weitz}(1998)}]{poulin.p:1998.a}
\bibinfo{author}{\bibfnamefont{P.}~\bibnamefont{Poulin}} \bibnamefont{and}
  \bibinfo{author}{\bibfnamefont{D.~A.} \bibnamefont{Weitz}},
  \bibinfo{journal}{Phys.\ Rev.\ E} \textbf{\bibinfo{volume}{57}},
  \bibinfo{pages}{626} (\bibinfo{year}{1998}).

\bibitem[{\citenamefont{Anderson et~al.}(2001)\citenamefont{Anderson,
  Terentjev, Meeker, Crain, and Poon}}]{anderson.vj:2001.a}
\bibinfo{author}{\bibfnamefont{V.~J.} \bibnamefont{Anderson}},
  \bibinfo{author}{\bibfnamefont{E.~M.} \bibnamefont{Terentjev}},
  \bibinfo{author}{\bibfnamefont{S.~P.} \bibnamefont{Meeker}},
  \bibinfo{author}{\bibfnamefont{J.}~\bibnamefont{Crain}}, \bibnamefont{and}
  \bibinfo{author}{\bibfnamefont{W.~C.~K.} \bibnamefont{Poon}},
  \bibinfo{journal}{Eur. Phys. J.} \textbf{\bibinfo{volume}{4}},
  \bibinfo{pages}{11} (\bibinfo{year}{2001}).

\bibitem[{\citenamefont{Anderson and Terentjev}(2001)}]{anderson.vj:2001.b}
\bibinfo{author}{\bibfnamefont{V.~J.} \bibnamefont{Anderson}} \bibnamefont{and}
  \bibinfo{author}{\bibfnamefont{E.~M.} \bibnamefont{Terentjev}},
  \bibinfo{journal}{Eur. Phys. J.} \textbf{\bibinfo{volume}{4}},
  \bibinfo{pages}{21} (\bibinfo{year}{2001}).

\bibitem[{\citenamefont{Meeker et~al.}(2000)\citenamefont{Meeker, Poon, Crain,
  and Terentjev}}]{meeker.sp:2000.a}
\bibinfo{author}{\bibfnamefont{S.~P.} \bibnamefont{Meeker}},
  \bibinfo{author}{\bibfnamefont{W.~C.~K.} \bibnamefont{Poon}},
  \bibinfo{author}{\bibfnamefont{J.}~\bibnamefont{Crain}}, \bibnamefont{and}
  \bibinfo{author}{\bibfnamefont{E.~M.} \bibnamefont{Terentjev}},
  \bibinfo{journal}{Phys.\ Rev.\ E} \textbf{\bibinfo{volume}{61}},
  \bibinfo{pages}{R6083} (\bibinfo{year}{2000}).

\bibitem[{\citenamefont{Russel et~al.}(1989)\citenamefont{Russel, Saville, and
  Schowalter}}]{russel.wb:1989.a}
\bibinfo{author}{\bibfnamefont{W.}~\bibnamefont{Russel}},
  \bibinfo{author}{\bibfnamefont{D.}~\bibnamefont{Saville}}, \bibnamefont{and}
  \bibinfo{author}{\bibfnamefont{W.}~\bibnamefont{Schowalter}},
  \emph{\bibinfo{title}{Colloidal dispersions}} (\bibinfo{publisher}{Cambridge
  University Press}, \bibinfo{address}{Cambridge}, \bibinfo{year}{1989}).

\bibitem[{\citenamefont{Brochard and de~Gennes}(1970)}]{brochard.f:1970.a}
\bibinfo{author}{\bibfnamefont{F.}~\bibnamefont{Brochard}} \bibnamefont{and}
  \bibinfo{author}{\bibfnamefont{P.~G.} \bibnamefont{de~Gennes}},
  \bibinfo{journal}{J. Phys} \textbf{\bibinfo{volume}{31}},
  \bibinfo{pages}{6911} (\bibinfo{year}{1970}).

\bibitem[{\citenamefont{Kuksenok et~al.}(1996)\citenamefont{Kuksenok, Ruhwandl,
  Shiyanovskii, and Terentjev}}]{kuksenok.ov:1996.a}
\bibinfo{author}{\bibfnamefont{O.~V.} \bibnamefont{Kuksenok}},
  \bibinfo{author}{\bibfnamefont{R.~W.} \bibnamefont{Ruhwandl}},
  \bibinfo{author}{\bibfnamefont{S.~V.} \bibnamefont{Shiyanovskii}},
  \bibnamefont{and} \bibinfo{author}{\bibfnamefont{E.~M.}
  \bibnamefont{Terentjev}}, \bibinfo{journal}{Phys.\ Rev.\ E}
  \textbf{\bibinfo{volume}{54}}, \bibinfo{pages}{5198} (\bibinfo{year}{1996}).

\bibitem[{\citenamefont{Shiyanovskii and
  Kuksenok}(1998)}]{shiyanovskii.sv:1998.a}
\bibinfo{author}{\bibfnamefont{S.~V.} \bibnamefont{Shiyanovskii}}
  \bibnamefont{and} \bibinfo{author}{\bibfnamefont{O.~V.}
  \bibnamefont{Kuksenok}}, \bibinfo{journal}{Mol.\ Cryst.\ Liq.\ Cryst.}
  \textbf{\bibinfo{volume}{321}}, \bibinfo{pages}{489} (\bibinfo{year}{1998}).

\bibitem[{\citenamefont{Mondain-Monval
  et~al.}(1999)\citenamefont{Mondain-Monval, Dedieu, Gulik-Krzywicki, and
  Poulin}}]{mondain-monval.o:1999.a}
\bibinfo{author}{\bibfnamefont{O.}~\bibnamefont{Mondain-Monval}},
  \bibinfo{author}{\bibfnamefont{J.~C.} \bibnamefont{Dedieu}},
  \bibinfo{author}{\bibfnamefont{T.}~\bibnamefont{Gulik-Krzywicki}},
  \bibnamefont{and} \bibinfo{author}{\bibfnamefont{P.}~\bibnamefont{Poulin}},
  \bibinfo{journal}{Euro.\ Phys.\ J. B} \textbf{\bibinfo{volume}{12}},
  \bibinfo{pages}{167} (\bibinfo{year}{1999}).

\bibitem[{\citenamefont{Ruhwandl and Terentjev}(1997)}]{ruhwandl.rw:1997.b}
\bibinfo{author}{\bibfnamefont{R.~W.} \bibnamefont{Ruhwandl}} \bibnamefont{and}
  \bibinfo{author}{\bibfnamefont{E.~M.} \bibnamefont{Terentjev}},
  \bibinfo{journal}{Phys.\ Rev.\ E} \textbf{\bibinfo{volume}{56}},
  \bibinfo{pages}{5561} (\bibinfo{year}{1997}).

\bibitem[{\citenamefont{Stark}(2001)}]{stark.h:2001.a}
\bibinfo{author}{\bibfnamefont{H.}~\bibnamefont{Stark}},
  \bibinfo{journal}{Physics Reports} \textbf{\bibinfo{volume}{351}},
  \bibinfo{pages}{387} (\bibinfo{year}{2001}).

\bibitem[{\citenamefont{Billeter and Pelcovits}(2000)}]{billeter.jl:2000.a}
\bibinfo{author}{\bibfnamefont{J.~L.} \bibnamefont{Billeter}} \bibnamefont{and}
  \bibinfo{author}{\bibfnamefont{R.~A.} \bibnamefont{Pelcovits}},
  \bibinfo{journal}{Phys.\ Rev.\ E} \textbf{\bibinfo{volume}{62}},
  \bibinfo{pages}{711} (\bibinfo{year}{2000}).

\bibitem[{\citenamefont{Andrienko et~al.}(2001)\citenamefont{Andrienko,
  Germano, and Allen}}]{andrienko.d:2001.b}
\bibinfo{author}{\bibfnamefont{D.}~\bibnamefont{Andrienko}},
  \bibinfo{author}{\bibfnamefont{G.}~\bibnamefont{Germano}}, \bibnamefont{and}
  \bibinfo{author}{\bibfnamefont{M.~P.} \bibnamefont{Allen}},
  \bibinfo{journal}{Phys.\ Rev.\ E} \textbf{\bibinfo{volume}{63}},
  \bibinfo{pages}{041701} (\bibinfo{year}{2001}).

\bibitem[{\citenamefont{Gu and Abbott}(2000)}]{gu.y:2001.a}
\bibinfo{author}{\bibfnamefont{Y.}~\bibnamefont{Gu}} \bibnamefont{and}
  \bibinfo{author}{\bibfnamefont{N.~L.} \bibnamefont{Abbott}},
  \bibinfo{journal}{Phys.\ Rev.\ Lett.} \textbf{\bibinfo{volume}{85}},
  \bibinfo{pages}{4719} (\bibinfo{year}{2000}).

\bibitem[{\citenamefont{Loudet and Poulin}(2001)}]{loudet.jc:2001.a}
\bibinfo{author}{\bibfnamefont{J.~C.} \bibnamefont{Loudet}} \bibnamefont{and}
  \bibinfo{author}{\bibfnamefont{P.}~\bibnamefont{Poulin}},
  \bibinfo{journal}{Phys.\ Rev.\ Lett.} \textbf{\bibinfo{volume}{87}},
  \bibinfo{pages}{165503} (\bibinfo{year}{2001}).

\bibitem[{\citenamefont{Burylov and Raikher}(1994)}]{burylov.sv:1994.a}
\bibinfo{author}{\bibfnamefont{S.~V.} \bibnamefont{Burylov}} \bibnamefont{and}
  \bibinfo{author}{\bibfnamefont{Y.~L.} \bibnamefont{Raikher}},
  \bibinfo{journal}{Phys.\ Rev.\ E} \textbf{\bibinfo{volume}{50}},
  \bibinfo{pages}{358} (\bibinfo{year}{1994}).

\bibitem[{\citenamefont{Stephen and Straley}(1974)}]{stephen.mj:1974.a}
\bibinfo{author}{\bibfnamefont{M.}~\bibnamefont{Stephen}} \bibnamefont{and}
  \bibinfo{author}{\bibfnamefont{J.}~\bibnamefont{Straley}},
  \bibinfo{journal}{Rev.\ Mod.\ Phys.} \textbf{\bibinfo{volume}{46}},
  \bibinfo{pages}{617} (\bibinfo{year}{1974}).

\bibitem[{\citenamefont{Roth et~al.}(2002)\citenamefont{Roth, van Roij,
  Andrienko, Mecke, and Dietrich}}]{roth.r:2001.a}
\bibinfo{author}{\bibfnamefont{R.}~\bibnamefont{Roth}},
  \bibinfo{author}{\bibfnamefont{R.}~\bibnamefont{van Roij}},
  \bibinfo{author}{\bibfnamefont{D.}~\bibnamefont{Andrienko}},
  \bibinfo{author}{\bibfnamefont{K.~R.} \bibnamefont{Mecke}}, \bibnamefont{and}
  \bibinfo{author}{\bibfnamefont{S.}~\bibnamefont{Dietrich}}
  (\bibinfo{year}{2002}), \bibinfo{note}{cond-mat/0202443}.

\bibitem[{\citenamefont{Berne and Pechukas}(1972)}]{berne.bj:1972.a}
\bibinfo{author}{\bibfnamefont{B.~J.} \bibnamefont{Berne}} \bibnamefont{and}
  \bibinfo{author}{\bibfnamefont{P.}~\bibnamefont{Pechukas}},
  \bibinfo{journal}{J. Chem.\ Phys.} \textbf{\bibinfo{volume}{56}},
  \bibinfo{pages}{4213} (\bibinfo{year}{1972}).

\bibitem[{\citenamefont{Gay and Berne}(1981)}]{gay.jg:1981.a}
\bibinfo{author}{\bibfnamefont{J.~G.} \bibnamefont{Gay}} \bibnamefont{and}
  \bibinfo{author}{\bibfnamefont{B.~J.} \bibnamefont{Berne}},
  \bibinfo{journal}{J. Chem.\ Phys.} \textbf{\bibinfo{volume}{74}},
  \bibinfo{pages}{3316} (\bibinfo{year}{1981}).

\bibitem[{\citenamefont{Lebwohl and Lasher}(1972)}]{lebwohl.pa:1972.a}
\bibinfo{author}{\bibfnamefont{P.~A.} \bibnamefont{Lebwohl}} \bibnamefont{and}
  \bibinfo{author}{\bibfnamefont{G.}~\bibnamefont{Lasher}},
  \bibinfo{journal}{Phys.\ Rev.\ A} \textbf{\bibinfo{volume}{6}},
  \bibinfo{pages}{426} (\bibinfo{year}{1972}).

\bibitem[{\citenamefont{Berggren et~al.}(1994)\citenamefont{Berggren, Zannoni,
  Chiccoli, Pasini, and Semeria}}]{berggren.e:1994.b}
\bibinfo{author}{\bibfnamefont{E.}~\bibnamefont{Berggren}},
  \bibinfo{author}{\bibfnamefont{C.}~\bibnamefont{Zannoni}},
  \bibinfo{author}{\bibfnamefont{C.}~\bibnamefont{Chiccoli}},
  \bibinfo{author}{\bibfnamefont{P.}~\bibnamefont{Pasini}}, \bibnamefont{and}
  \bibinfo{author}{\bibfnamefont{F.}~\bibnamefont{Semeria}},
  \bibinfo{journal}{Phys.\ Rev.\ E} \textbf{\bibinfo{volume}{50}},
  \bibinfo{pages}{2929} (\bibinfo{year}{1994}).

\bibitem[{\citenamefont{Metropolis and Ulam}(1949)}]{metropolis.n:1949.a}
\bibinfo{author}{\bibfnamefont{N.}~\bibnamefont{Metropolis}} \bibnamefont{and}
  \bibinfo{author}{\bibfnamefont{S.}~\bibnamefont{Ulam}}, \bibinfo{journal}{J.
  Amer.\ Stat.\ Ass.} \textbf{\bibinfo{volume}{44}}, \bibinfo{pages}{335}
  (\bibinfo{year}{1949}).

\bibitem[{\citenamefont{Fabbri and Zannoni}(1986)}]{fabbri.u:1986.a}
\bibinfo{author}{\bibfnamefont{U.}~\bibnamefont{Fabbri}} \bibnamefont{and}
  \bibinfo{author}{\bibfnamefont{C.}~\bibnamefont{Zannoni}},
  \bibinfo{journal}{Molec.\ Phys.} \textbf{\bibinfo{volume}{58}},
  \bibinfo{pages}{763} (\bibinfo{year}{1986}).

\bibitem[{\citenamefont{Barker and Watts}(1969)}]{barker.ja:1969.a}
\bibinfo{author}{\bibfnamefont{J.~A.} \bibnamefont{Barker}} \bibnamefont{and}
  \bibinfo{author}{\bibfnamefont{R.~O.} \bibnamefont{Watts}},
  \bibinfo{journal}{Chem.\ Phys.\ Lett.} \textbf{\bibinfo{volume}{3}},
  \bibinfo{pages}{144} (\bibinfo{year}{1969}).

\bibitem[{\citenamefont{Priezjev and Pelcovits}(2000)}]{priezjev.n:2000.a}
\bibinfo{author}{\bibfnamefont{N.}~\bibnamefont{Priezjev}} \bibnamefont{and}
  \bibinfo{author}{\bibfnamefont{R.~A.} \bibnamefont{Pelcovits}},
  \bibinfo{journal}{Phys.\ Rev.\ E} \textbf{\bibinfo{volume}{62}},
  \bibinfo{pages}{6734} (\bibinfo{year}{2000}).

\bibitem[{\citenamefont{Braun and Viney}(2001)}]{braun.fn:2001.a}
\bibinfo{author}{\bibfnamefont{F.}~\bibnamefont{Braun}} \bibnamefont{and}
  \bibinfo{author}{\bibfnamefont{C.}~\bibnamefont{Viney}},
  \bibinfo{journal}{Phys.\ Rev.\ E} \textbf{\bibinfo{volume}{63}},
  \bibinfo{pages}{031708} (\bibinfo{year}{2001}).

\bibitem[{\citenamefont{Schopohl and Sluckin}(1987)}]{schopohl.n:1987.a}
\bibinfo{author}{\bibfnamefont{N.}~\bibnamefont{Schopohl}} \bibnamefont{and}
  \bibinfo{author}{\bibfnamefont{T.~J.} \bibnamefont{Sluckin}},
  \bibinfo{journal}{Phys.\ Rev.\ Lett.} \textbf{\bibinfo{volume}{59}},
  \bibinfo{pages}{2582} (\bibinfo{year}{1987}).

\bibitem[{\citenamefont{Sonnet et~al.}(1995)\citenamefont{Sonnet, Kilian, and
  Hess}}]{sonnet.a:1995.a}
\bibinfo{author}{\bibfnamefont{A.}~\bibnamefont{Sonnet}},
  \bibinfo{author}{\bibfnamefont{A.}~\bibnamefont{Kilian}}, \bibnamefont{and}
  \bibinfo{author}{\bibfnamefont{S.}~\bibnamefont{Hess}},
  \bibinfo{journal}{Phys.\ Rev.\ E} \textbf{\bibinfo{volume}{52}},
  \bibinfo{pages}{718} (\bibinfo{year}{1995}).

\bibitem[{\citenamefont{Germano}(2001)}]{germano.preprint}
\bibinfo{author}{\bibfnamefont{G.}~\bibnamefont{Germano}}
  (\bibinfo{year}{2001}), \bibinfo{note}{in preparation}.

\end{thebibliography}

\end{document}